\documentclass[twocolumn,prl,aps,epsf]{revtex4}

    \newif \ifpdf
    \ifx \pdfoutput\undefined
    \pdffalse 
    \else
    \pdfoutput=1 
    \pdftrue
    \fi

    \ifpdf
    \usepackage[pdftex]{graphicx}
    \else
    \usepackage{graphicx}
    \fi

\begin{document}

 \ifpdf
    \DeclareGraphicsExtensions{.pdf, .jpg, .tif}
    \else
    \DeclareGraphicsExtensions{.eps, .jpg}
    \fi
\draft


\title{Microwave resonance of the reentrant insulating quantum Hall phases in the 1st excited Landau Level}
\author{R. M. Lewis$^{1,2}$, Yong P.\ Chen$^{1,2}$, L. W. Engel$^1$, D. C. Tsui$^2$, L. N. Pfeiffer$^{3}$, and K. W. West$^{3}$
}

\address{$^{1}$NHMFL, Florida State University, Tallahassee, FL 32310, USA\\
$^{2}$Dept.\ of Electrical Engineering, Princeton University, Princeton, NJ 08544\\
$^{3}$Bell Laboratories, Lucent Technologies, Murray Hill, NJ 07974}

\date{\today}
\begin{abstract}We present measurements of the real diagonal microwave conductivity of the reentrant insulating quantum Hall phases in the first excited Landau level at temperatures below 50 mK.  A resonance is detected around filling factor $\nu=2.58$ and weaker frequency dependence is seen at $\nu=2.42$ and 2.28.  These measurements are consistent with the formation of a bubble phase crystal centered around these $\nu$ at very low temperatures.
\end{abstract}

\maketitle    
\vskip1pc

Clean two--dimensional electron systems (2DES) show a plethora of phases at low temperatures ($T$) when subjected to perpendicular magnetic fields ($B$).  The most well known of these are the integer\cite{kvk} and fractional\cite{tsui} quantum Hall effects (IQHE and FQHE).  Both are identified by quantized Hall resistances, $R_{xy}=h/\nu e^2 $ and the simultaneous vanishing of the diagonal resistance ($R_{xx}$) when the filling factor $\nu=n h/eB$ coincides with a gap in the single or many-particle density of states.  In higher Landau levels (LL), $N \ge 2$, where $N$ is the LL index, vanishing minima in $R_{xx}$ are observed\cite{mlilly,rrdu} near $\nu=4+1/4$, 4+3/4, and several subsequent odd quarter $\nu$.  However, at these $\nu$, $R_{xy}$ is quantized to the value of the {\it adjacent} integer plateau.  Therefore these states are said to exhibit a reentrant integer quantum Hall effect (RIQHE)\cite{cooper_iv}.  Theory \cite{fogler,moessner,kunyang,yoshioka} predicts a crystal phase of the 2DES with triangular lattice symmetry and {\it two or more} electrons per lattice site---pinned by disorder and hence insulating---to occur at approximately these $\nu$.  This crystal has been dubbed the bubble phase (BP).  Experiments have subsequently shown non-linear current--voltage characteristics \cite{cooper_iv}, a microwave resonance\cite{rlewis}, and narrow band noise\cite{cooper_noise}, in the range of $\nu$ centered around $\nu=4+1/4$.  Still more recent work\cite{rlewis_coex} has looked at the transition between the BP around $\nu=4 +1/4$ and another solid phase---presumably a Wigner crystal\cite{wcreviews,ychen,rlewis2} of carriers in the uppermost LL---close to $\nu=4$.

The microwave measurements \cite{rlewis,rlewis_coex}, which are similar to those we will discuss here, provide compelling evidence that the RIQHE is due to the formation of a pinned electron solid such as the BP.  These measurements detect a sharp resonance in the real diagonal conductivity (Re[$\sigma_{xx}(f)$]) when the 2DES forms an electron solid. The observed resonant frequency, $f_{pk}$, is considerably lower, in terms of energy, than the temperature at which the resonance first appears, $f_{pk} << k_B T/ h$, indicating that the resonance is not due to the ionization of individual carriers trapped in potential defects.  Furthermore, this resonance is strikingly similar to the resonance which occurs in the insulating phase of 2DES\cite{lloydssc,peidewc} that terminates the FQHE at high $B$, $\nu <1/5$, where the 2DES is thought to form a Wigner crystal\cite{wcreviews}.   The resonances in both regimes (RIQHE and $\nu < 1/5$ insulating phase) are interpreted as pinning modes\cite{fukuyama_lee,chitra} of crystal domains oscillating within the disorder potential.

Recently, Eisenstein {\it et al.} \cite{eisenstein} and Xia {\it et al.} \cite{xia_n_wei} have presented measurements  of RIQHE's in the $N=1$ LL which appear only at $T \lesssim 50$ mK.  These RIQHE's are centered at partial filling factors $\nu^*\approx 0.28$, 0.42, 0.56, and 0.70 where $\nu^*= \nu- [\nu]$ and $[\nu]$ is the greatest integer less than $\nu$.  The initial theories\cite{fogler,moessner} of charge density waves in higher LL (which predicted the stripe and BP for $\nu> 4$) did not consider the N=1 LL.  Also, the short range softening of the effective electron--electron potential, which gives rise to the BP in higher LL, is only weakly present in the N=1 LL and absent in the N=0 LL.  However, more recent theories\cite{yoshioka2,goerbig,goerbig2} do predict a BP and perhaps a stripe phase in the first excited LL.  

    In this paper, we present measurements of Re$[\sigma_{xx}]$ between $\nu=2$ and 3.  At $T \approx 35$ mK, we show that a resonance occurs in the frequency ($f$) dependence of Re$[\sigma_{xx}]$ in a range of $\nu$ around 2.58.  An enhancement of Re$[\sigma_{xx}]$ at $\nu \approx 2.42$ is also observed at low $f$.  These features are not present for $T \ge 55$ mK.   The resonance occurs at frequency, $f \sim 125$ MHz, and is coincident with the RIQHE phase found previously \cite{eisenstein,xia_n_wei} in the N=1 LL on either side of $\nu=5/2$.  In light of our earlier observations\cite{rlewis} in higher LL's, this resonance is naturally interpreted as due to the pinning mode of the BP in the N=1 LL at $\nu=2.57$.

The sample, grown by molecular beam epitaxy, is a 300 \AA\ GaAs/AlGaAs quantum well of density n=3.0$\times 10^{11}\  {\rm cm^{-2}}$ and mobility $\mu=2.4 \times 10^7 \ {\rm cm^2 \  V^{-1} \ s^{-1}}$ at 300 mK.  The quantum well is approximately 2000\AA \ below the surface.  A coplanar waveguide \cite{wen} (CPW) is patterned onto the sample surface.  The CPW consists of a driven centerline of length $l$ which is separated from symmetrically placed ground planes by slots of width $w$.  At high $f$ and small $|\sigma_{xx}|$, the conductivity of the 2DES attenuates power transmission along the CPW and Re$[\sigma_{xx}]= -\frac{w}{2lZ_0} ln(P_t/P_0)$, where $P_t$ is the transmitted power\cite{lloydprl93}.  In the absence of 2DES, the CPW has impedance $Z_0=50\ \Omega$ and transmits power $P_0$.  The CPW pattern is shown in the inset to Fig.\ 1 and has $l=28$ mm and $w=30\ \mu$m.  The dark regions are the metal films of the CPW.  Measurements were carried out in the low power limit which was determined by reducing microwave power until no further change in Re$[\sigma_{xx}]$ was observed. 

\begin{figure}[tb]
\begin{center}
\includegraphics[width=8.5cm]{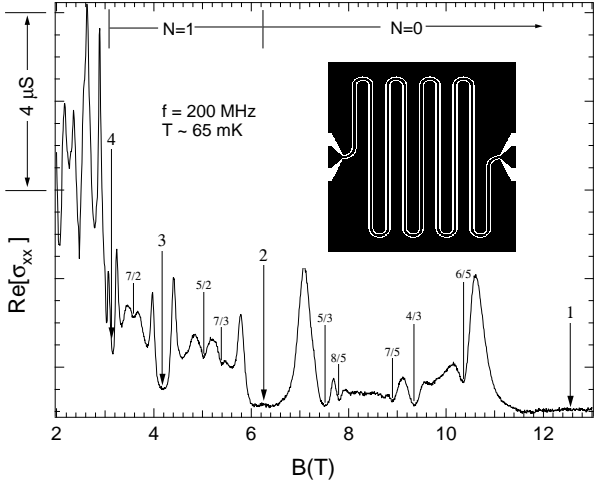}
\caption{ The real part of the diagonal conductivity, Re$[\sigma_{xx}]$ versus $B$ between 2 and 13 T. {\bf Inset} is the coplanar waveguide pattern in which dark regions are the metal films.}
\label{fig.1}
\end{center}
\end{figure}

In Fig.\ 1 we show Re$[\sigma_{xx}]$ measured, between 2 and 13 T at 200 MHz and $T \approx 65$ mK.  The broad minima centered around $B=12.55$, 6.26, 4.18, and 3.14 T are due to the IQHE states at $\nu=1$, 2, 3, and 4.  Between $\nu=1$ and 2, sharp minima indicate FQHE states in the N=0 LL at $\nu=6/5$, 4/3, 7/5, 8/5, and 5/3.  The N=1 LL also shows FQHE states.  For instance, at $B=5.38$, 5.03, and 3.58 T, dips due to the $\nu=7/3$, the 5/2, and 7/2 FQHE are apparent.  For $\nu >4$ (N=2 LL), sharp peaks appear in Re$[\sigma_{xx}]$ at $B \approx 2.89$, 2.63, 2.36, and 2.17 T, or roughly at $\nu \approx 4 +1/4$, $4+3/4$, $5+1/4$, and $5 +3/4$ where DC measurements \cite{cooper_iv, mlilly,rrdu} observe the RIQHE.  These peaks are due to a resonance at $f \approx 300$ MHz, which was interpreted\cite{rlewis} as due to the pinning mode\cite{fukuyama_lee} of the BP.

\begin{figure}[tb]
\begin{center}
\includegraphics[width=8.8cm]{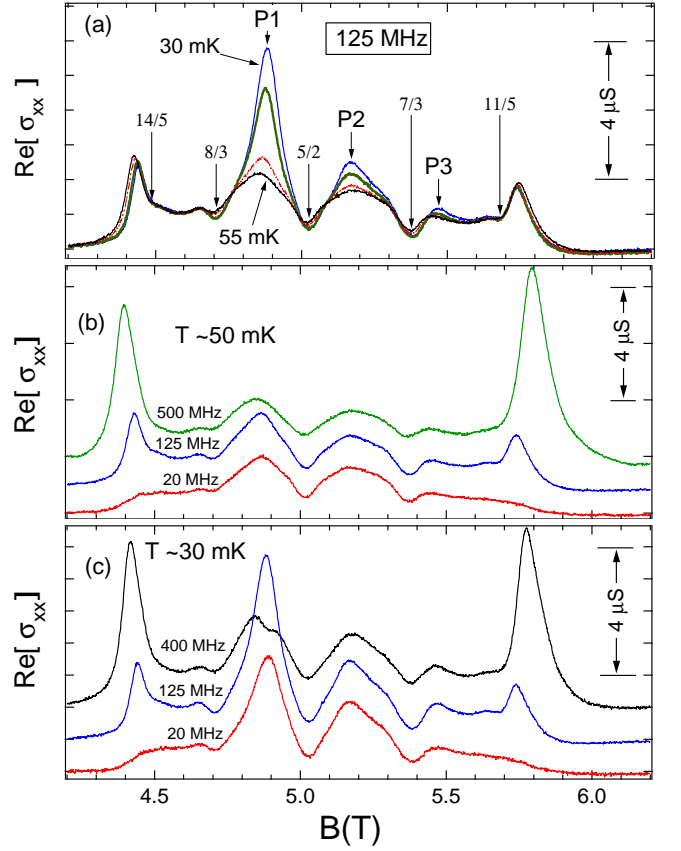}
\caption{ 
{\bf (a)} Re[$\sigma_{xx}$] vs $B$ between 4.2 and 6.2 T at frequency $f=125$ MHz for temperatures $T \approx$ 30, 40, 50, and 55 mK.   {\bf (b)}  Re[$\sigma_{xx}$] vs $B$ at $f=20$, 125, and 500 MHz  with $T \approx 50$ mK offset 1 $\mu$S from each other.  {\bf (c)}, Re[$\sigma_{xx}$] vs $B$ at $f=20$, 125, and 400 MHz, all at $T =30$ mK offset for clarity.
}
\label{fig.2}
\end{center}
\end{figure}

In Fig.\ 2, we focus on the range of $B$ between 4.2 and 6.2 T where $\nu$ lies within the N=1 spin--up LL.  In Fig.\ 2a, we show Re$[\sigma_{xx}]$ at $f=125$ MHz for $T$ approximately at 30, 40, 50 and 55 mK.  The most striking aspects of these data are the dramatic increases of the peaks in Re$[\sigma_{xx}]$  at 4.88 T (marked as P1) and at 5.17 T (marked as P2) with decreasing $T$.  P1 occurs at $\nu=2.57$ and  P2 at $\nu=2.42$.  At P1, Re$[\sigma_{xx}]$ increases by $\sim 4\ \mu$S as $T$ decreases from 55 mK to 30 mK.  Similarly, at P2, Re$[\sigma_{xx}]$ increases by about 1 $\mu$S over the same $T$ range.  Some $T$ dependence is also evident in the peak at $B= 5.47$ T ($\nu=2.28$), labelled as P3.  

Fig.\ 2b shows Re$[\sigma_{xx}]$ vs $B$ measured at $f=20$, 125, and 500 MHz at $T \approx 50$ mK.  At this $T$,  P1 and P2, show little dependence on $f$. The traces are offset 1 $\mu$S from each other. 

In Fig.\ 2c, we plot Re$ [\sigma_{xx}]$ vs $B$ for  $f=20$, 125, and 400 MHz, at $T \approx 30 $ mK offset by 1 $\mu$S each for clarity.  At P1, Re$[\sigma_{xx}]$ increases by roughly 2 $\mu$S as $f$ goes from 20 to 125 MHz where we use $\nu=2$ at 6.2 T as the reference point.  However, Re$ [\sigma_{xx}]$ decreases by  about 3 $\mu$S as $f$ changes from 125 to 400 MHz, indicating that Re$ [\sigma_{xx}]$ vs $f$ has a resonance near $f=125$ MHz.  At P2, an increase of 0.2 $\mu$S is seen between 20 and 125 MHz and a decrease of 0.2 $\mu$S occurs between 125 and 400 MHz.  The variation of Re$ [\sigma_{xx}]$ with $f$ at P3, however, is too weak to quantify.

Dependence of Re$[\sigma_{xx}]$ on $f$ is also seen close to the IQHE minima in Fig.\ 2b and 2c at $B= 4.43$ and 5.75 T.  Detailed discussions of these $f$ dependences can be found in Refs. \cite{ychen} and \cite{rlewis2} and is attributed to the formation of a pinned electron solid in the uppermost partially filled LL near integer $\nu$.

\begin{figure}[tb!]
\begin{center}
\includegraphics[width=8.8cm]{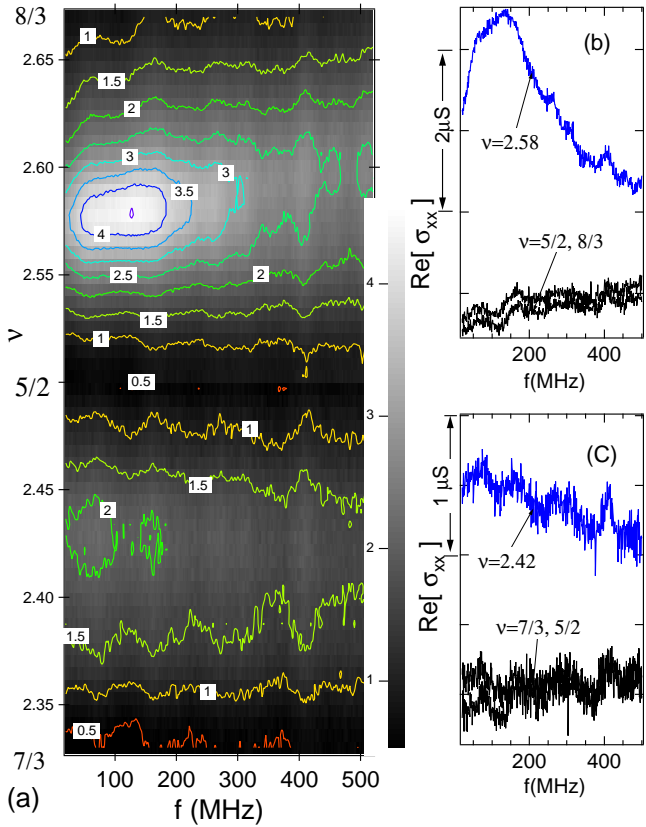}
\caption{{\bf (a)}The real diagonal conductivity, Re$[\sigma_{xx}]$ vs frequency ($f$) in grayscale, between $\nu=7/3$ and 8/3 at $T \approx 35$ mK. Contours of constant conductivity are superimposed at intervals of 0.5 $\mu$S. {\bf (b)} Re$[\sigma_{xx}]$ at $\nu=2.58$ compared with $\nu=5/2$ and 8/3. {\bf(c)}  Re$[\sigma_{xx}]$ at $\nu=2.42$ compared with $\nu=7/3$ and 5/2.}
\label{fig.3}
\end{center}
\end{figure}

In Fig.\ 3a, we plot Re$[\sigma_{xx}]$ in grayscale for $f$ in the 20 to 520 MHz range and $\nu$ between $\nu=$ 7/3 and 8/3.  Contours of constant Re$[\sigma_{xx}$] are superimposed at 0.5 $\mu$S intervals.  Re$[\sigma_{xx}$] shows dark bands at the 7/3, 5/2, and 8/3 FQHE states indicating minima, as expected from the $B$ dependence shown in Fig.\ 2.  Lighter regions in between the FQHE states indicate increased  Re$[\sigma_{xx}$].  At $\nu \approx 2.58$, a resonance in Re$[\sigma_{xx}$]  vs $f$ is seen which peaks at$f \approx 120$ MHz and is encircled by the contour lines.  This resonance is sharpest and between $\nu=2.57$ ad 2.59, but can be observed for $\nu$ between 2.54 and 2.61.  In Fig.\ 3b, a spectrum of Re$[\sigma_{xx}]$ vs. $f$ at $\nu=2.58$ is shown which exhibits a broad resonance at $f \approx 120$ MHz.  For comparison, Re$[\sigma_{xx}]$ spectra taken at $\nu= 5/2$ and 8/3 are also shown in Fig.\ 3b and are featureless.

In the lower half of Fig.\ 3a, some increase in  Re$[\sigma_{xx}$] is observed between $\nu=7/3$ and 5/2.  But, the contour lines run nearly parallel to each other as only non-resonant frequency dependence is seen.  A weak upturn in Re$[\sigma_{xx}]$ is seen between $\nu \approx 2.37$ and $\nu \approx 2.46$ and is most pronounced  at $\nu=2.42$.  Fig.\ 3c shows that Re$[\sigma_{xx}]$ vs $f$ measured at  $\nu=2.42$ exhibits a steady {\it decrease} as $f$ {\it increases}.  This dependence is quite different from the gentle increase in Re$[\sigma_{xx}]$ vs $f$ measured at  $\nu=7/3$ and 5/2.  At this sensitive scale, small oscillations are visible in the data which are due to weak reflections between the sample and the room temperature amplifier.  The roughly 1 $\mu$S difference in Re$[\sigma_{xx}]$ at 500 MHz is due to the increase in nonresonant background conductivity between the FQHE minima at 7/3 and 5/2 and is also seen in the data in Fig.\ 2.   The error in our determination of $\nu$  in Fig.\ 3 is $\pm 0.01$.

The phenomenology of these N=1 LL RIQHE phases is similar to what has been observed in higher LLs.  The dc measurements \cite{eisenstein,xia_n_wei} show that around $\nu^*=$ 0.28, 0.42, 0.58, and 0.70 insulating phases appear in the uppermost partially filled LL for $T \lesssim 50$ mK.  The resonance in Re$[\sigma_{xx}]$ shown in Fig.\ 3b strongly suggests that the insulating phase centered around $\nu=2.58$ is collective in nature---most likely an electron solid---because $f_{pk} \ll k_B T/h$.  The predictions of theory \cite{yoshioka2,goerbig,goerbig2} are that this solid is a two electron per lattice site BP.  Therefore, we naturally interpret the resonance we find at $\nu \approx 2.58$ as due to a pinned BP.  Furthermore, near $\nu \approx 2.42$, we find an enhancement of Re$[\sigma_{xx}]$ at low $f$ which is manifestly different from spectra taken at the 5/2 and 7/3 FQHE states. Rather, the spectra at $\nu=2.42$ is more similar to the resonant behavior at $\nu=2.58$ and hence suggestive of pinned collective insulating behavior.

Recent density matrix renormalization group calculations by Shibata and Yoshioka \cite{yoshioka2} find a unidirectional charge density wave (stripe) phase for $\nu$ between the 11/5 and 7/3 FQHE states and also between the 7/3 and 5/2  FQHE.  Those authors suggest however, that disorder may favor the BP over the stripe phase within these ranges of $\nu$.  Mean field theory by Goerbig {\it et al.} \cite{goerbig,goerbig2} finds only a BP in these same ranges of $\nu$ and predicts a transition from the 7/3 FQHE to the BP at $\nu^*= 0.36$ close to what we observe.  The theories \cite{yoshioka2,goerbig,goerbig2} invoke particle--hole symmetry for $\nu^*=0.5$ to 1.  In contrast, our measurements down to $T=35$ mK show that the BP centered around $\nu=2.57$ develops at higher $T$ than the BP centered at $\nu=2.42$.  Strangely, this differs from observations of the BP in N=2 and higher LL, where the electron branch ($\nu^* <0.5$) always persists to higher $T$ than the hole branch ($\nu^*>0.5$) of the BP.  The development of the 12/5 FQHE state\cite{xia_n_wei} at extremely low $T<25$ mK is likely responsible. 
 
In summary, we have studied the $f$ dependence of the RIQHE phases in the N=1 LL between $\nu=2$ and 3 for $T \le 55$ mK.
We find a resonance in Re$[\sigma_{xx}]$ vs $f$ for $\nu=2.58$, where $f_{pk} = 120 \pm 15$ MHz measured at 35 mK.  The resonance appears for $\nu$ between 2.54 and 2.61 and is extraordinarily similar to measurements of the BP in the N=2 and higher LLs\cite{rlewis}.  Somewhat surprisingly, a resonance has not developed at $T \approx 35$ mK near $\nu=2.42$, but an upturn in Re[$\sigma_{xx}(f)$] at low $f$ is observed between $\nu=2.37$ and 2.46 at $35$ mK.  The ranges of $\nu$ where these dependences are seen coincide with RIQHE phases reported in Refs. \cite{eisenstein} and \cite{xia_n_wei} and with theoretical predictions of the BP\cite{yoshioka2,goerbig,goerbig2} in the N=1 LL .  We interpret these data as consistent with formation of an electron solid for $\nu$ between 2.54 and 2.61, presumably a BP, at $T\lesssim 50$ mK.

We thank Kun Yang and Herb Fertig for stimulating discussions. The measurements were performed at the NHMFL with financial support from the AFOSR and the NHMFL in--house research program.  The NHMFL is operated under NSF grant  DMR--0084173 with support from the state of Florida.

\end{document}